# Metabolic scaling, life history, and the equal fitness paradigm

Joseph R. Burger

Department of Biology, University of Kentucky, Lexington, KY 40506 USA

Email: robbieburger@uky.edu

## Summary

Natural selection has produced an extraordinary diversity of life histories spanning many orders of magnitude in body size, vital rates, and biological times. In general, big and cold organisms grow and reproduce slowly and live long lives; small and warm organisms grow and reproduce quickly and live short lives. The Metabolic Theory of Ecology (MTE) predicts equal and opposite scaling exponents of mass-specific biological rates (e.g., respiration, growth, and reproduction) and times (e.g., development, lifespan, and generation) as a function of size. However, empirical support for these predictions varies depending on trait and taxon. Here I: 1) provide background and mixed support for the quarter-power scaling exponents for life history rates and times predicted by MTE, 2) discuss possible explanations, including effects of natural selection on taxonomic and functional groups, and inadequate data for life history traits, 3) briefly summarize the Equal Fitness Paradigm (EFP) as a unifying theory of bioenergetics, life history and demography that does not depend on any particular allometric scalings, and 4) discuss ramifications of the EFP for other biological phenomena, including physiological performance metrics and trophic energetics of ecosystems. I draw mostly from my knowledge of mammals, yet in many cases the mammalian examples can be generalized to other organisms. I end with prospects for further evaluating and extending the EFP.

## Introduction

Metabolism, the process of acquiring and transforming energy and materials to perform biological work, is fundamental to the organization of living systems and the pace of life. It profoundly influences every aspect of life, from the structure and function of organisms to the dynamics of populations and ecosystems, and the origin and maintenance of biodiversity. Across biodiversity—from microbes to plants and animals; and scales and levels of organization from cells to the biosphere—metabolism fuels life.

In this chapter, I first provide a brief history of metabolic scaling. Then I review the mixed support for quarter-power scalings of life history rates and times. I then discuss possible explanations for this discrepancy between theory and data. I then briefly summarize the Equal Fitness Paradigm (EFP: Brown et al. 2018, 2022; 2024; Burger et al. 2019; 2021; 2023) as a unifying theoretical framework to account for the diversity of life history traits. I end by discussing ramifications of the EFP for aspects of biological organization, from life history trade offs, population dynamics and trophic energetics of ecosystems.

## Metabolic theory

Decades of research quantified how physiological traits vary with body size and temperature including Rubner (1883), Kleiber (1932), Brody (1945) and Hemmingsen (1960). A seminal finding that respiration rates of mammals and birds increase non-linearly as the ¾ power of body mass known as "Kleiber's rule" led to a flurry of empirical studies synthesized in several books (Peters (1986), McMahon and Bonner (1983), Calder (1984) and Schmidt-Nielsen (1984)). Initial data confirmed that the exponents are often close to "quarter powers" – and significantly different from "third powers" that would be expected from standard geometric scaling (i.e., of length, surface area and volume). Moreover, scalings can be grouped into categories based on "the principle of similitude" and dimensional analysis, so that whole-organism rates scale as mass, $m^{3/4}$, mass-specific rates as $m^{-1/4}$, and biological times as the reciprocals of the rates as $\sim m^{1/4}$. But there was no unifying explanation for these scalings.

In the late 1990s another round of studies was inspired by the West, Brown, Enquist model (West et al. 1997) that attributed the quarter powers to the fractal-like architecture of the biological transport systems that supply and remove metabolites. The Metabolic Theory of Ecology (MTE; Brown et al. 2004) subsequently proposed a general equation for how rates and times scale as a function of size and temperature taking the mathematical form

$$Y = Y_o m^\alpha e^{\frac{E_a}{kt}} \qquad (1)$$

where $Y$ is the value of a life history time (e.g., development or generation time), whole organism and mass-specific rates (e.g., respiration, heart rate, biomass production), $Y_o$ is a normalization coefficient, $m$ is body mass, $\alpha$ is the mass-scaling exponent, $e$ is the base of the natural logarithm, $E_a$ is an "activation energy" which gives the temperature dependence, $k$ is Boltzmann's constant, and $t$ is the temperature in Kelvin (Gillooly 2001; Brown et al. 2004; Sibly et al. 2012a).

MTE sought to explain the mechanistic processes underlying these scaling relations by parameterizing mass-specific metabolic rates, $B$, and many other rates scaling with $\alpha \approx$ -1/4 and $E_a \approx$ -0.65, eV ≈ $Q_{10}$ ~2.5, and lifespans, $L$, and many other biological times scaling with $\alpha \approx$ 1/4 and $E_a \approx$ 0.65 eV (Gillooly 2001; Brown et al. 2004; Sibly et al. 2012a). In birds and mammals with relatively constant body temperatures (~37-40 °C), eqn 1 can be simplified to body size allometries where the mass-specific rates scale as $B = B_0 M^{-1/4}$ and times as $L = L_0 M^{1/4}$.

Much of the research on allometric scaling, metabolism and life history has focused on the tradeoff between metabolic rate and lifespan and corresponding "Rate of Living" hypotheses (RoL: Pearl 1928; Stearns 1992; Roff 2002; Speakman 2005). The RoL posits that the total energy expenditure on respiration and production or total number of heartbeats in a lifetime is constant irrespective of size:

$$Y = BL = Y_0 M^\alpha = B_0 M^{-1/4} = L_0 M^{1/4} = Y_0 M^0 \tag{2}$$

A corollary is the so-called pace of life or fast-slow continuum, where small organisms live fast and die young, whereas large ones live slowly and die old. Implicit in eqn 2 is that all animals age at the same rate, independent of body size (and temperature). Recently, however, empirical patterns, underlying mechanisms, and theoretical contributions have failed to support the generality of both the empirical quarter-power exponents and the mechanistic fractal hypothesis (Recent reviews in Glazier 2022; Harrison *et al.* 2022; Glazier 2024). Next, I review the empirical data with mixed support for the MTE.

At the largest scales of geographic space, evolutionary time, and biological organization there are grounds to both support and question whether any universal theory of biological scaling and metabolism applies across all organisms (Glazier 2022). Across the broad spectrum of taxa and body sizes there is a qualitative and statistically significant tradeoff between speed of living and duration of life: unicellular protists live fast and giant mammals and trees live slow (Figure 1; Hatton et al. 2019; Burger et al. 2021). Figure 1 might be taken to imply that approximately quarter-power scaling is universal across ~22 orders of magnitude variation in size from unicellular organisms to giant trees and whales. Upon closer examination, however, this very general qualitative pattern across the tree of life (Figure 1) obscures considerable systematic and quantitative variation between and within taxonomic and functional groups that span narrower ranges of body sizes.

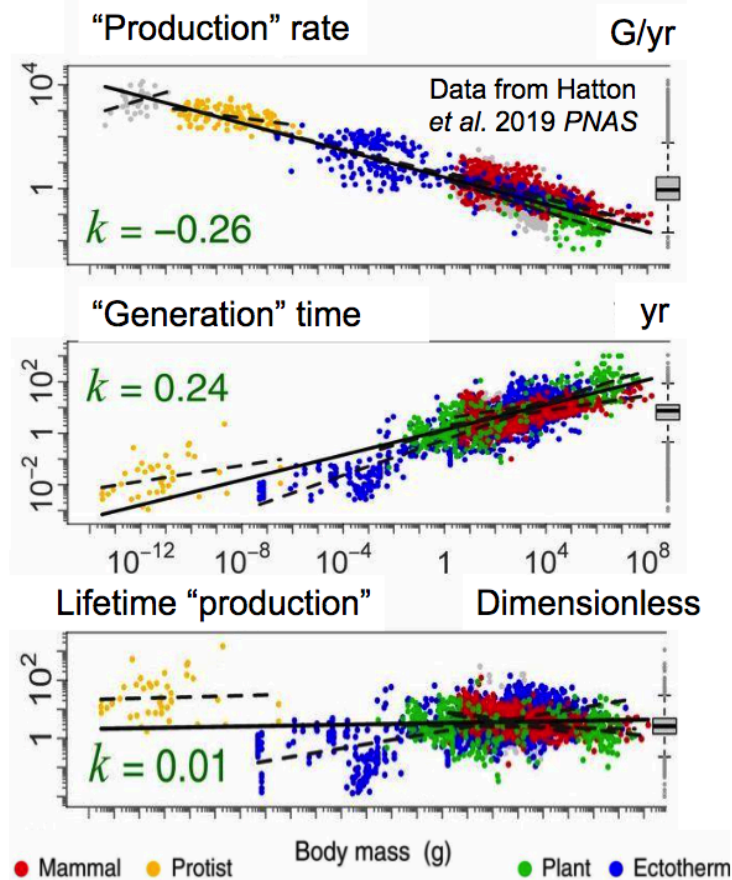

**Figure 1. The allometry of A) "mass-specific production rates", B) "Generation time" and C) "lifetime production" for species spanning >20 orders of magnitude variation in body size (After Hatton et al. 2019). However, note the substantial variation between and within the different groups**.

Brown & Sibly (2006) and Sibly & Brown (2007) showed that life histories of mammals exhibit two primary axes of variation: body size and lifestyle (Figure 2). Bigger mammals have slower mass-specific rates of respiratory metabolism and biomass production than smaller ones. Importantly, however, among species of similar body size, rates of mass-specific production vary by ~2 orders of magnitude. Sibly & Brown (2006, 2007) attributed this residual variation to a "lifestyle" axis orthogonal to size (Figures 2 & 3). For example, bats and primates have lower production rates and live longer than most mammals of similar size (Figure 3). Ungulates have relatively high production rates and short lifespans for their size. Rodents and carnivores exhibit wide variation, but most species are intermediates, representing "the norm" within the wide bounds for mammals. In part because examples of such "lifestyle" axes are widespread, theoretical predictions of quarter-power allometric scalings are not supported by studies of mammals and other taxonomic, phylogenetic and functional groups.

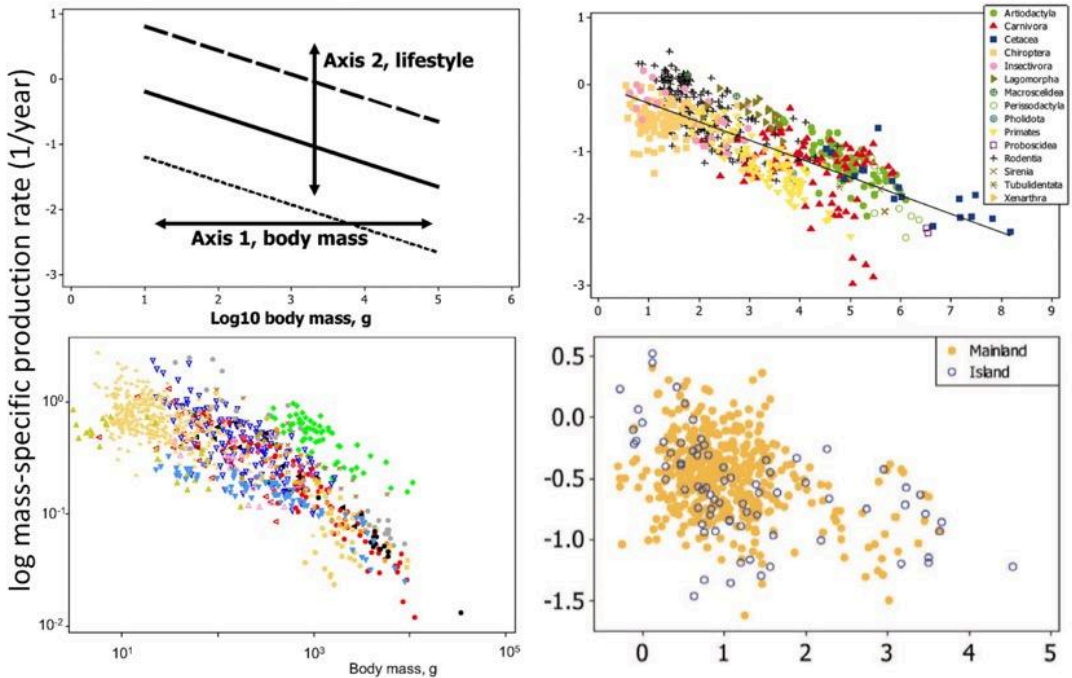

**Figure 2 illustrates the allometry of mass-specific production rates: A) a conceptual diagram showing primary (body size) and secondary (lifestyle) life history axes. Larger animals have slower production rates, while species above the line exhibit faster rates than expected for their size, B) eutherian mammals (After Sibly & Brown 2007), C) birds (After Sibly et al. 2012b), and D) mainland and island lizard species (After Meiri et al. 2012). Mass-specific production rates per year are calculated as: $\frac{m_w LB}{m_a}$ where $m_w$ is mass of the offspring size at birth, $L$ is litter size, $B$ is the number of reproductive bouts annually. Exponents differ from the -¼ scalings predicted by allometric and metabolic theories.**

The regression analyses in Figure 2 do not support α = - ¼ predictions, although the variation in production rates generally scales negative with size. Birds show wide variation in production rates with body size with a general curvilinear decreasing pattern (Sibly et al. 2012b). The "lifestyle" axis in lizard production has even more variation, yet generally decreasing with large size with a possible upper constraint (Meiri et al. 2012). In mammals, among species of similar body size, generation times vary by about one order of magnitude (Sibly and Brown 2007; Hamilton et al. 2011) and by at least two orders of magnitude in insects (Brown et al. 2022; Harrison et al. 2022) see also (Glazier 2022; Glazier 2024).

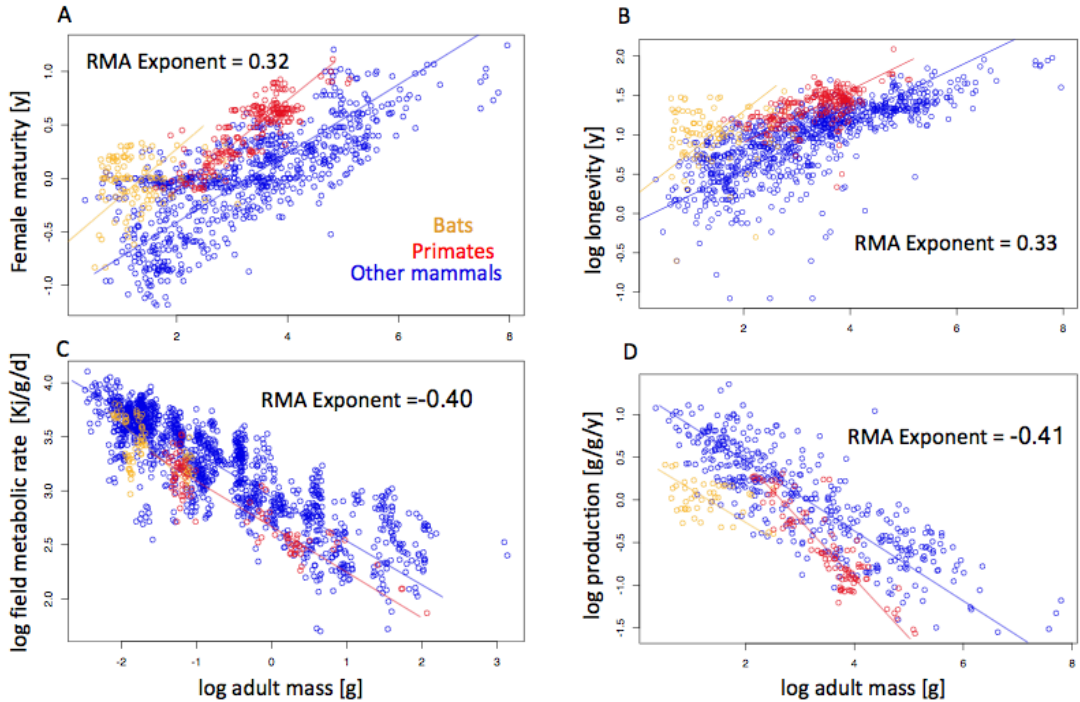

**Figure 3 shows examples of scalings of some life history and metabolic rates and times with body size in mammals using Reduced Major Axis (RMA) regression. Note the scaling exponents differ from MTE predictions of +-¼. Yet, bigger organisms have slower mass-specific rates and slower life history times than small organisms. Bats and primates are highlighted to show how multiple 'lifestyles' that reduce mortality are associated with an orthogonal axis of slower rates (below line) and longer times (above line). Data from Hudson et al. (2013) and Myhrvold et al. (2015).**

The unicellular microbes – at the small end of the size spectrum – show an interesting departure from theory (Figure 4; DeLong et al. 2010; Okie 2012; Lynch et al. 2022). DeLong et al. (2010) found that mass-specific metabolic and production rates scale positively with size in prokaryotes, with an exponent of 0.73, whereas protists and metazoans show ~ -1/4 scaling as predicted from traditional theory. There appears to be additional variation in scaling relations due to adaptation to different environments. Okie (2012) showed that peak production rates vary among microbes in response to temperature and pH gradients, reflecting unique adaptations to their abiotic environments. At the other end of the size spectrum, Brown et al. (2022) proposed that the orders-of-magnitude differences among of mammals and insects of the same body size in mass-specific production rates and generation times are due largely to natural selection to fit intrinsic biological times of life histories into the geochronological cycles of extrinsic environments. These geochronological cycles may explain much of the variation and departures from ¼ power expectations. The great diversity of "annual" temperate zone plants and animals with one-year generation times are more emblematic examples.

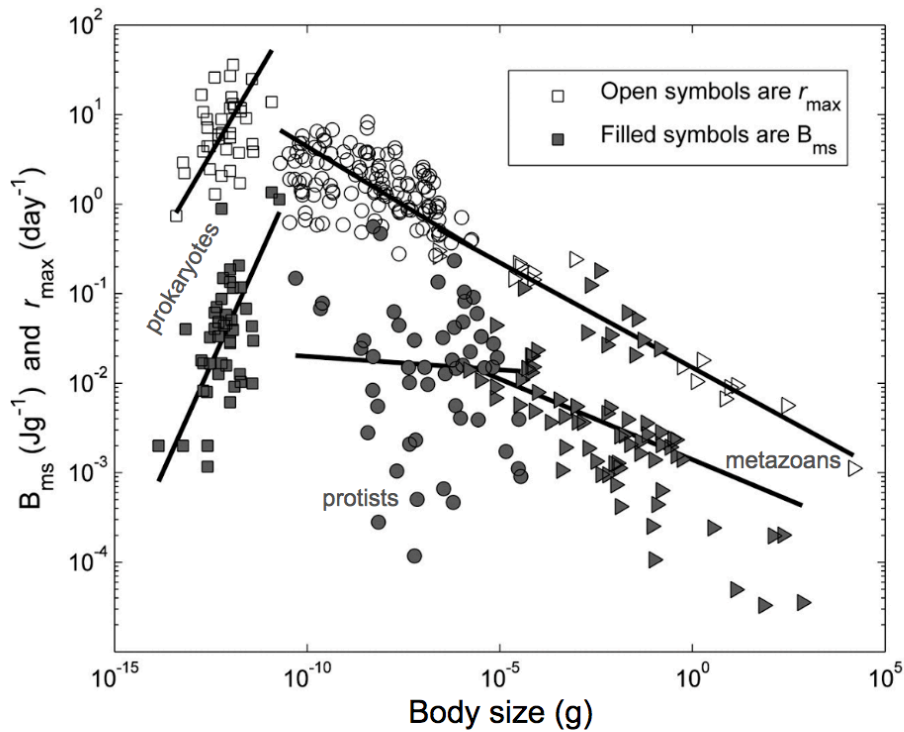

**Figure 4 shows the scaling of maximum population growth rate ($r_{max}$, open symbols) and mass-specific metabolic rate ($B_{ms}$, filled symbols) with body mass across a diversity of prokaryotes (squares), protists (circles) and metazoans (triangles). The exponents using Reduced Major Axis (RMA) are 0.73 for prokaryotes, −0.26 for protists, and −0.23 for metazoans. The authors conclude that because the scalings of $r_{max}$ and $B_{ms}$ within groups are statistically indistinguishable from parallel, the results support the hypothesis that metabolic rate fuels biomass production (after DeLong et al. 2010).**

## The equal fitness paradigm

*It is the theory which decides what we can observe* (Einstein)

Theory informs what parameters to measure and how to collect and analyze the data. The recently formulated Equal Fitness Paradigm (EFP) is an alternative to standard allometric and metabolic scaling theory because it: i) does not depend on ¼ power or any specific allometric scaling constants; and ii) enforces mass and energy balance and demographic steady state consistent with the laws of physics and demography, and iii) provides a theoretical framework that defines the relevant traits and informs how to measure them.

When considering what variables to measure and evaluate theories of allometric scaling and metabolic ecology, it is useful to construct a mass-energy balance diagram (Figure 5) to quantify stocks and flows of energy and biomass over a life cycle similar to a Dynamic Energy Budget following Kooijman (2010). Animals, plants, and microbes assimilate energy ($A$) from the environment and allocated between metabolic respiration ($R$) burned off as heat and biomass production ($P$):

$$A = R + P \quad (3).$$

The metabolic cost of maintenance, activity, and producing offspring is included in respiration; growth, storage, and offspring investment in biomass are included in production. Metabolic and biomass production rates have been investigated extensively in mammals, but assimilation rates rarely.

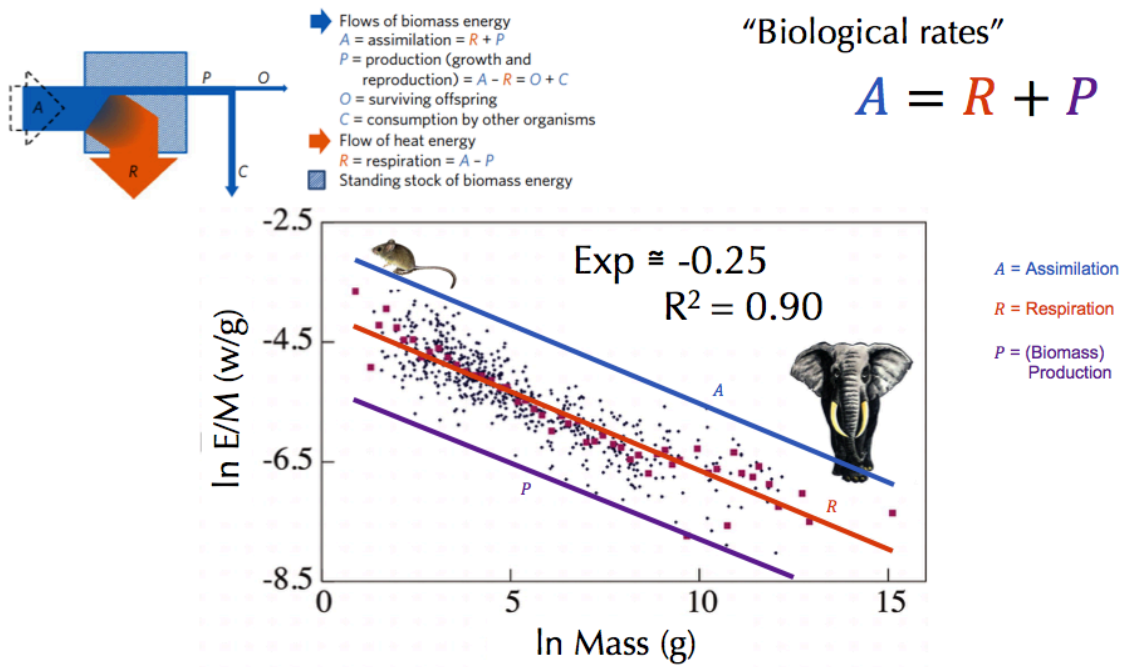

**Figure 5 shows empirical scalings of mass-specific metabolic rate ($R$) with body size in mass from Savage et al. (2007). A mass-energy balance diagram is depicted in the upper left insert. Organisms assimilate energy from the environment ($A$) and allocate it between respiration ($R$): catabolic metabolism to synthesize ATP to perform the biological work of living, ultimately dissipated as heat; and ii) production ($P$): net new biomass of individual offspring growth ($H$) and parental reproductive investment ($I$). So $A = P + R$ and $P = H + I$. According to the principle of similitude, we would expect $R$, $P$ and $A$ to be parallel with similar exponents and offset constants in agreement with the provided equation.**

The earliest presentation of the EFP built on RoL, MTE and allometric scaling relations (Brown et al. 2018). It assumed that the relevant production rate, $P_{ind}$, was the rate for an individual organism and used data such as in Figures 1-3 to parameterize the variables and evaluate the predictions. Subsequent development of the EFP parameterized production in terms of the biomass of the *cohort*, $P_{coh}$, the rate of biomass production in offspring until they die including those that grow independently of the parent (Burger et al. 2019, 2021, 2023; Brown et al. 2022, 2024). This requires the biomass accounting to include not only investments of mature parents in gametes and offspring nutrition, but also investments of offspring in ontogenetic growth between independence from parental inputs and reproductive maturity – so importantly, including the biomass of offspring that died without reproducing.

The result is the equation of the EFP
$$M = P_{coh}GF \tag{4}$$
where energetic fitness, $M$ is the biomass invested in surviving offspring, $P_{coh}$ is the rate of production of the entire cohort of offspring, $G$ is generation time, and $F$ is the fraction of cohort production that is passed on to surviving offspring. Eq 4 can be parameterized in units of either energy or mass, e.g., $M$ has units of g/g per generation ($G$), $P_{coh}$ units g/g/y, $G$ units of years, and $F$ is the dimensionless fraction of $M$ that is passed on to surviving offspring.

The EFP provides an alternative, yet nonexclusive, life history theory that addresses many of the same phenomena of allometry, metabolic ecology and evolutionary biology. It is based on two fundamental biophysical constraints on all living things: (i) mass-energy balance, where assimilated energy is allocated to respiration to pay the energy costs of living, and biomass production to leave surviving offspring; and (ii) demographic balance, where, regardless of the number of offspring produced, only one survives to replace the parent in the next generation.

These allocations of energy and biomass – between respiration to synthesize ATP and perform work, and production for ontogenetic growth and parental investment – are required to maintain a steady-state population and avoid extinction. The EFP expresses traditional components of Darwinian fitness – survival and reproduction – in the biological currency of energy and biomass consistent with the laws of energy conservation and demographic balance.

## Important differences among theories

The EFP is generally consistent with MTE and standard scaling relations, but universal scaling parameters are not required. While body size and temperature may correlate with $P_{coh}$, and $G$, the EFP does not depend on any particular allometry or temperature dependence. The EFP differs from earlier theories in at least four important respects:

1) The time scale is one generation.—The relevant time scale in which parents produce offspring to replace themselves is one generation $G$, the time from zygote to zygote or the lifespan of an average individual that survived from birth to reproductive maturity.

This allows various life history times such as time to independence from parent, developmental time, age and maturity to normalize by generation time using dimensionless fractions. This can be calculated from demographic life tables, where $R_o$ is the net replacement rate, $s_x$ and $l_x$ are survival and fecundity as a function of age.

2) The biophysical currency is production.—From the perspective of fitness, the most important rate is not metabolic rate of individual organism respiration, $R$, the rate of synthesis of ATP to power the biological work of maintenance, growth and reproduction. The assimilated biomass expended on $R$ is dissipated back into the environment as heat, carbon dioxide and water; none is passed on to the next generation as fitness. The relevant rate of allocation of assimilated biomass to fitness is the rate of cohort production, $P_{coh}$, the mass-specific rate of biomass produced of all of the offspring in a cohort per generation, including parental investment in gametes and nutrition given to offspring prior to independence (e.g., weaning in mammals) and independent (e.g., post-weaning) ontogenetic growth of individuals.

3) The unit of allocation is a cohort.—Assimilated biomass is allocated to respiration plus production in accord with energy balance (Figure 5). At steady state, when the population is stationary and parents exactly replace themselves with offspring in one generation, all biomass energy taken up from the environment in assimilation is returned to the environment as heat in respiration plus biomass of dead offspring (i.e., the "ecosystem tax"; Burger et al. 2023). The total production is the cohort

$$A_{coh} = P_{coh} + R_{coh} \tag{5}$$

where $P_{coh}$ is the mass-specific rate of biomass produced by the cohort of offspring up until death, including the one "survivor" that ultimately dies after reproducing to replace its parent. The central equation of the EFP (eqn. 4) highlights the importance of the parameter, $F$, the fraction of cohort biomass produced per generation that is passed to surviving offspring and actually realized as fitness.

4) The growth and mortality of offspring are included.—The original version of the EFP (Brown et al. 2018) followed earlier life history theories (Stearns 1992; Roff 2002; Charnov et al. 2007) in parameterizing production as investments in gametes and offspring nutrition prior to independence from parental inputs. Consequently these earlier formulations violated energy balance by not incorporating postnatal mortality and growth of offspring, including those that died before maturing and reproducing. This problem is corrected by parameterizing in terms of $P_{coh}$, the production of the entire cohort, and $F$, the efficiency of reproduction: the fraction of production passed on to surviving offspring:

$$F = \frac{2m_A}{m_{tot}} = \frac{2m_A}{2m_A + m_{pre}} \tag{6}$$

where $2m_A$ is the body mass of the two offspring that survive to replace the parents, $m_{tot}$ is the total biomass produced by the entire cohort in one generation, and $m_{pre}$ is the biomass of offspring that grow to varying extents and die without leaving surviving offspring; $m_{tot} = \sum_{x=0}^{x=G} m_x d_x$, where $m_x$ is the body mass and $d_x$ is the number of offspring dying at age $x$. The complementary fraction, $C = 1 - F$, is the "ecosystem tax" (Burger et al. 2023), the fraction of cohort production lost to pre-reproductive mortality, and $m_{tot}$ is the total biomass recycled in the ecosystem per generation. Theory and initial data suggests values of $C$ can vary from ~0.50 in highly social mammals and microbes with higher parental investment to >0.99 in highly fecund fishes and marine invertebrates (Burger et al. 2023). This framework can be extended to predict the ecosystem consequences of the varying tax when populations are not in steady-state, linking life histories to trophic energetics.

## New data opportunities

The above differences highlight the need for better data. It might seem that the increasing availability of big data for many traits and taxa of organisms would have led to rapid progress in biological scaling and metabolic ecology. However, many of the large databases used in comparative life history studies are not carefully documented and standardized (Duncil, Cook, Burger unpublished). The interpretation of the data also depends on the magnitude of variation and the scale of analysis. Differences by a factor of 2 and even 10 are seemingly insignificant across life where the variation in body size spans many orders of magnitude (Figure 1), but it is much more apparent and potentially important within a taxonomic and functional group with a much narrower size range (Figures 2-4).

For example, "lifespan" is commonly reported in comparative databases and is fundamentally different from the generation time of the EFP. Average lifespans of mammals and birds in free-living populations in the wild can be an order of magnitude shorter than maximum lifespans of individuals kept under favorable conditions in captivity. Lifespans of populations in the wild, rather than being characterized by unique species-specific values, can vary substantially, depending on environment (e.g. climate and habitat) and demography (e.g., whether the population is increasing, stationary or decreasing; human managed or naturally controlled).

A case in point. The widely used Amniote database (Myhrvold et al. 2015) includes data for lifespan of white-tailed deer as 21.6 (y). Using data from wild populations in steady state, (Burger *et al.* 2019) calculated average generation time to be close to 2 years. Use of such inflated lifespans to estimate lifetime fecundity or production rates in deer and many organisms can result in substantial overestimates. New databases will be required because current comparative life history datasets are insufficient. To parameterize the EFP and accurately calculate the efficiency of reproduction, $F$, [$C = 1 - F$, the ecosystem tax of Burger et al. (2023)] requires a metabolic life table (MLT) that quantifies growth, survival and fecundity over a complete life cycle including iteroparous and semelparous species (Van Valen 1975; Brett 1983, 1986; Burger *et al.* 2020, 2021; Brown et al. 2022, 2024).

## Conclusions and road forward

Recent reviews (Sibly et al. 2012; Glazier 2022) suggest that after more than a century of accumulating data and theory, there is still no consensus – no widely accepted general model or theory for how and why the basic structures and functions of living things vary with body size. The EFP moves beyond these historical studies of scaling to re-formulate and re-evaluate how the components of energy, biomass and life history determine fitness. While size-scaling and temperature-dependent relations are pervasive across the entire tree of life (Hemingson 1960, Brown et al. 2004; Hatton et al. 2019), these offer only very imprecise predictors of many aspects of individual structure and function, ecosystem energetics and biodiversity.

Consider the following example. Suppose we want to predict the metabolic rate, lifespan and lifetime fecundity of an impala, another artiodactyl about the same size as a white-tailed deer. We could potentially plug in to increasingly specific published allometric equations for all animals, all mammals, non-volant placental mammals, or artiodactyls. Such fitted scaling relations inevitably trade off generality for precision, and not all are currently available for all three life history traits at all four levels of taxonomic resolution. Consequently, the predicted values will vary considerably. Empirical values to evaluate the predicted ones are not currently available for all categories, and some that do exist are of questionable utility due to issues of standardization of definition of traits and methods of measurement (see above). The most precise prediction requires specific equations for the most restricted taxonomic category (artiodactyls), and the most precise empirical test requires a complete metabolic life table for impala detailed. Furthermore, there will still be some magnitude of mismatch between predicted and observed values calling for additional explanation.

Although the EFP does not assume any explicit scaling relations, it highlights three important conceptual issues that are directly relevant. First, by enforcing mass and energy balance and demographic steady state consistent with the laws of physics, the EFP specifies how to collect and analyze new data of life history traits. The EFP emphasizes that natural selection only acts indirectly on body size (or any other trait's) scaling exponents and coefficients (Brown et al. 2024). Selection affects the evolution of these traits by operating directly on the three parameters, $P_{coh}$, $G$, and $F$ so as to maximize the product $P_{coh}GF$ (eq 4). A complete accurate metabolic life table is required to evaluate the EFP empirically and account mechanistically for the evolution of life history traits.

Second, the EFP offers an alternative explanation for the pervasive tradeoff between respiration rate and lifespan (RoL) that stimulated much of the work on biological scaling. Rather than incorporating current metabolic scaling theory and assuming any specific body size-dependent parameters, the EFP is based on the formulation of energetic fitness in eq. 4. It parameterized the variation in life histories among three life history traits that are measurable for all organisms: rate of cohort biomass production, ($P_{coh}$), generation time ($G$) and efficiency of production ($F$).

Finally, the EFP additionally highlights the importance of $F$, the fraction of biomass in surviving offspring and, $C = 1 - F$, the ecosystem tax, in connecting individual energy budgets to trophic energetics (Burger et al. 2023). Interestingly, $F$ and $C$, which quantify mortality, are two life history traits that do not scale significantly with body size across the diversity of life (Burger et al. 2023; Brown et al. 2024). Rather, the the ecosystem tax, $C$, varies greatly—from ~40% in the smallest organisms (unicellular microbes) and in larger animals that provide extensive parental care (sharks, birds and mammals)—to >99% in large species that produce enormous numbers of minute offspring (large trees, teleost fishes and marine invertebrates). Today's industrial economy runs on fossil fuels that are taxes paid by dead organisms millions of years ago. The EFP provides a synthetic and universal theory to understand the generation and maintenance of biodiversity across space and time in accordance with mass-energy balance. It may provide additional insights into practical questions in human ecology and sustainability (e.g., Burger et al. 2012, 2017).